# Circular dichroism mode splitting and bounds to its enhancement with cavity-plasmon-polaritons

*Denis G. Baranov\*, Battulga Munkhbat, Nils Odebo Länk, Ruggero Verre, Mikael Käll, and Timur Shegai*

Department of Physics, Chalmers University of Technology, 412 96 Göteborg, Sweden

*email: denisb@chalmers.se

**Abstract**: The ability to differentiate chiral molecules of different handedness is of great importance for chemical and life sciences. Since most of the relevant chiral molecules have their chiral transitions in the UV region, detecting their circular dichroism (CD) signal is associated with practical experimental challenges of performing optical measurements in that spectral range. To address this problem, here, we study the possibility of shifting CD signal of a model chiral medium by reaching the strong coupling regime with an optical microcavity. Specifically, we show that by strongly coupling chiral plasmonic nanoparticles to a non-chiral Fabry-Pérot microcavity one can imprint the hybrid mode splitting, the hallmark of strongly coupled systems, on the CD spectrum of the coupled system and thereby effectively shift the chiral resonance of the model system to lower energies. We first predict the effect using analytical transfer-matrix method as well as numerical finite-difference time-domain (FDTD) simulations. Furthermore, we confirm the validity of theoretical predictions in a proof-of-principle experiment involving chiral plasmonic nanoparticles coupled to a Fabry-Pérot microcavity.

**Introduction**

An object is said to be chiral, if it cannot be superimposed with its mirror image by a series of rotations and displacements. The two mirrored versions in this case are called the left and right *enantiomers*. Geometrical chirality has fundamental implications on the electromagnetic response of chiral systems and materials. Any geometrically chiral absorptive object, being it a molecule or an optical nanoantenna, will exhibit circular dichroism (CD), that is, non-equal extinction of left and right circularly polarized (CP) light, in the absence of any magneto-optical activity [1,2].

Geometrical chirality is ubiquitous in nature – it is encountered on many different length scales ranging from geometrical shapes of various living organisms to protein and DNA molecules. When acting on a biological receptor, opposite enantiomers of the same molecule may induce different response, such as odor or taste [3]. It is therefore important to have efficient ways of separation of different enantiomers of chiral biomolecules and detect their conformal changes [1,4,5]. This ability is even more pivotal for pharmaceutics, where a wrong enantiomer in a drug may be highly poisonous [6]. The phenomenon of CD provides one possible way of optical discrimination between different enantiomers that otherwise look identical [7,8]. The ability to enhance CD from chiral molecules is therefore of enormous importance for modern biotechnology.

Enhancement of light-matter interactions is typically achieved by using various kinds of optical cavities and resonators. It is therefore natural to expect that interfacing chiral molecules with such optical resonators would lead to enhanced CD signal. Over the years, many attempts to realize these expectations have been made, including usage of Fabry-Pérot microcavities [9,10], hotspots of plasmonic nanoantennas [11–17], or chiral metasurfaces producing superchiral light [7,18]. The effect of the cavity leads to a local enhancement of *optical chirality* of incident light [19,20] – a quantity determining the degree of asymmetry between the interaction strengths of chiral light with left and right enantiomers [21,22]. Since chiral transitions of relevant organic or biomolecules often occur in the UV spectral range [11,14], CD measurements have to face experimental challenges associated with measurements in this spectral range.

Here, we show that it is in principle possible to shift the optical response of a generic chiral scatterer, such as a chiral molecule or a chiral nanoparticle, to a different frequency region by using the concept of strong coupling [23–25]. When the interaction rate of the material resonance with a cavity mode exceeds their corresponding decay rates, the system enters the strong coupling regime, characterized by the emergence of mode splitting and formation of intermixed light-matter eigenstates – polaritons. We show that not only scattering and absorption, but also CD of a coupled system exhibits mode splitting in the strong coupling domain. At the same time, we reveal fundamental limitations on the enhancement of CD signal in planar resonators imposed by circularly polarized standing waves and suggest ways to overcome this limit. We verify this effect with FDTD simulations and a proof-of-principle experiment utilizing chiral plasmonic nanoparticles as prototypical chiral scatterers interacting with a Fabry-Pérot microcavity.

**Analytical calculations**

The system under study is schematically presented in Fig. 1 – it is formed by an array of chiral scatterers, such as chiral molecules, positioned inside a planar optical cavity. The chiral scatterers are expected to couple with the vacuum field of the Fabry-Pérot cavity mode, leading to the emergence of cavity-plasmon-polaritons, and giving rise to mode splitting in transmission and reflection spectra[26–28]. Due to the scatterers' inherent chirality, the system as a whole has to exhibit CD in transmitted signal. Owing to the formation of polariton eigenstates, one may expect that not only transmission and reflection spectra, but also the CD spectra of the initial chiral system will also exhibit a similar mode splitting.

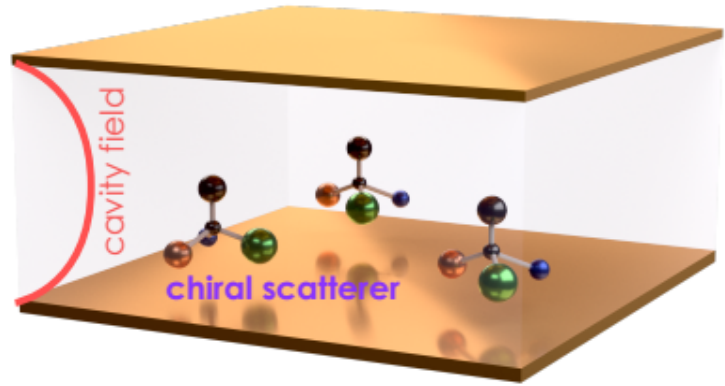

**Fig. 1**. Sketch of the system under study: chiral scatterers are placed inside a planar cavity.

As a starting point of the analysis, we model the chiral coupled system as a planar multilayer structure and calculate its optical response with the use of the transfer-matrix method (TMM, see Appendix A). The test structure consists of a bi-isotropic chiral layer placed between two gold mirrors that create a Fabry-Pérot cavity, as sketched in Fig. 2(a). The constitutive relations for a general non-magnetic bi-isotropic material in SI units read [29] $\begin{pmatrix} \mathbf{D} \\ \mathbf{B} \end{pmatrix} = \begin{pmatrix} \varepsilon\varepsilon_0 & -i\kappa/c \\ i\kappa/c & \mu_0 \end{pmatrix} \begin{pmatrix} \mathbf{E} \\ \mathbf{H} \end{pmatrix}$, where $\varepsilon_0$ and $\mu_0$ are the vacuum permittivity and permeability, $c$ is the speed of light, $\varepsilon$ is the relative permittivity, and $\kappa$ is the Pasteur parameter that couples the electric and magnetic fields. We model the chiral layer with a Lorentzian permittivity $\varepsilon = 1 + f_0 \frac{\omega_0^2}{\omega_0^2 - \omega^2 - i\gamma\omega}$ and the corresponding dispersive Pasteur parameter $\kappa = \kappa_0 \frac{\omega_0^2}{\omega_0^2 - \omega^2 - i\gamma\omega}$, where $\omega_0$ is the resonance frequency of the chiral transition, $\gamma$ is its linewidth the, $f_0$ is the oscillator strength, and $\kappa_0$ is the amplitude of the Pasteur parameter [29].

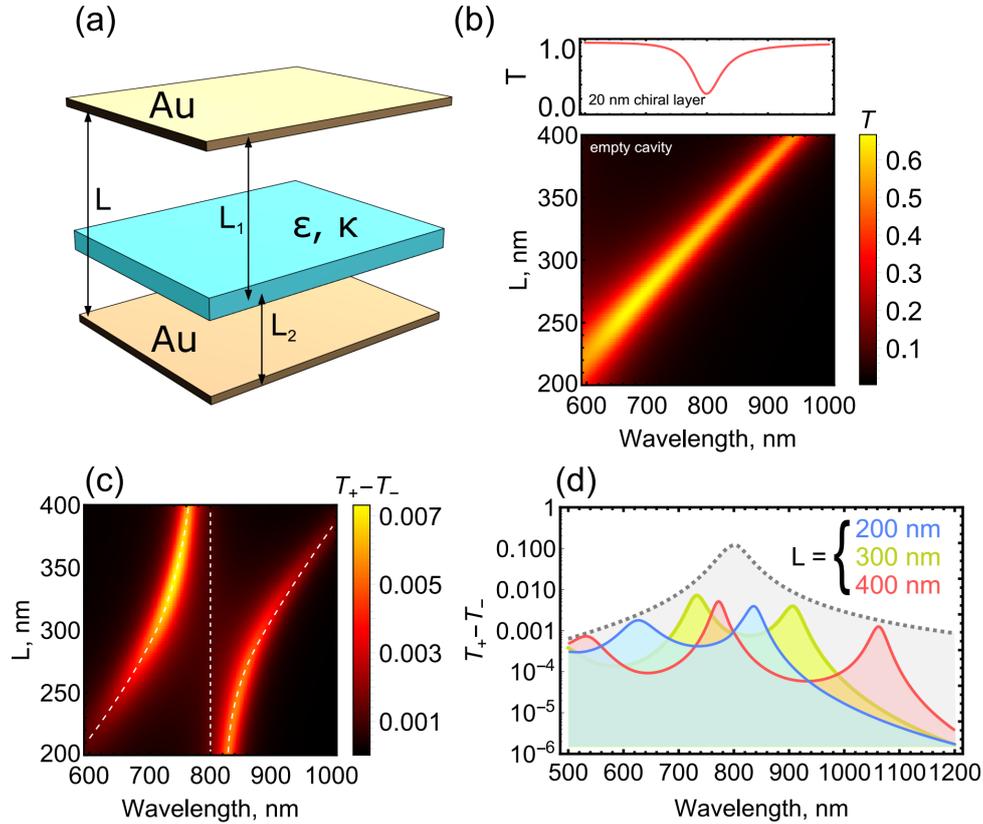

**Fig. 2.** Transfer-matrix calculations of a chiral multilayer system. (a) Sketch of the system studied with the TMM method: a bi-isotropic chiral layer characterized by a resonant permittivity $\varepsilon$ and Pasteur parameter $\kappa$ placed between two gold mirrors. (b) Transmission spectrum of a 20 nm thick bi-isotropic chiral layer in free space, and transmission spectra of an empty cavity as a function of its thickness. (c) CD spectra of the coupled system with the chiral film placed *in the middle* of the cavity as the function of the cavity thickness. (d) CD spectra of the 20 nm thick chiral film placed in the middle of the cavity of various thicknesses (solid curves) compared to that of the same film in vacuum (dashed grey). Note logarithmic scale in y-axis

Fig. 2(b) shows transmission spectra of a bare cavity formed by two 20 nm thick gold mirrors as a function of the cavity thickness $L$, and that of a 20 nm thick bi-isotropic chiral layer in free space for $\omega_0 = 1.55$ eV, $\gamma = 50$ meV, $f_0 = 0.5$, and $\kappa_0 = 0.03$. Owing to the Lorentzian dielectric response, the bi-isotropic layer is expected to strongly couple with the cavity mode and exhibit mode splitting in scattered (transmitted) signal. We then calculate CD spectra of the same chiral layer placed in the middle of the cavity as a function of the cavity thickness, Fig. 2(c). Calculated map of CD spectra clearly demonstrates expected anti-crossing. However, when we compare these CD spectra to that of the bi-isotropic layer in vacuum, we find that CD from the layer in vacuum at all wavelengths exceeds that of the layer inside the cavity, Fig. 2(d). We note here (and everywhere in this manuscript, unless stated otherwise) that CD was calculated as difference in transmission between two different CP incident beams, $T_+ - T_-$.

Noteworthy, such suppression of CD occurs despite enhanced EM field inside the cavity at the resonance, Fig. 3(a), and strong interaction between the cavity mode and the chiral layer manifested in the anti-crossing. The reason for this limitation becomes apparent when we look at the chirality density $C = -\frac{\varepsilon_0 \omega}{2} \text{Im}(\mathbf{E}^* \cdot \mathbf{H})$ induced by an incident CP wave inside the cavity. As shown in Fig. 3(b), chirality density normalized by that of the incident CP wave is always less than unity. Since the asymmetry of light-matter interaction and CD magnitude is linked to the chirality density [22], CD produced by a chiral layer inside a planar FP cavity is always smaller than the free space value. In fact, this limitation of chirality density is fundamental to any isotropic passive planar cavity (i.e., without gain).

Fig. 3(c) illustrates the nature of this limitation: let us consider a CP wave of a certain handedness (and carrying the corresponding chirality density $C_0 = -\frac{\varepsilon_0 \omega}{2} \text{Im}(\mathbf{E}_0^* \cdot \mathbf{H}_0)$) and intensity $I_0 = \frac{1}{2} \text{Re}(\mathbf{E}_0 \times \mathbf{H}_0)$ incident onto a planar cavity; let us assume for concreteness $C_0 > 0$. The cavity mirrors may have arbitrary composition; we will be interested in the empty cavity region between the mirrors where a chiral layer can be positioned later. As the wave experiences a series of reflections inside the cavity, the intensity $I_i$ of each consecutive wave can only decrease. At the same time, the handedness of the travelling wave, which determines the sign of the chirality density, is *reversed* upon each reflection [29]. Furthermore, since $C$ is a bi-linear function of fields, chirality density of each consecutive wave in the empty region between the mirrors is $C_i = (-1)^{i-1}(I_i/I_0)C_0$. Total chirality density in this field distribution is $C = -\frac{\varepsilon_0 \omega}{2} \text{Im}((\mathbf{E}_1 + \mathbf{E}_2 + \cdots + \mathbf{E}_i + \cdots)^* \cdot (\mathbf{H}_1 + \mathbf{H}_2 + \cdots + \mathbf{H}_i + \cdots))$, where $\mathbf{E}_i$ and $\mathbf{H}_i$ are the electric and magnetic field amplitudes of *i*-th circularly polarized wave. By direct calculations, one can easily verify that terms $\text{Im}(\mathbf{E}_i^* \cdot \mathbf{H}_j + \mathbf{E}_j^* \cdot \mathbf{H}_i)$ vanish for $i \neq j$, so the total chirality density reduces to the sum of chirality densities of individual waves:

$$C = C_1 + C_2 + \cdots + C_i + \cdots.$$

This is a sign-alternating sum, such that $|C_{i+1}| < |C_i|$; therefore, it converges, and clearly $|C| < C_0$. This proves that chirality density inside any planar passive isotropic cavity is smaller than that of the incident CP wave, which imposes bounds on CD produced by a chiral layer placed inside the cavity. Of course, this bound can be lifted by incorporating gain medium, which would allow transmitted waves to have larger intensity and chirality densities. However, from the energy point of view it is equivalent to increase of the incident CP wave intensity in a passive system.

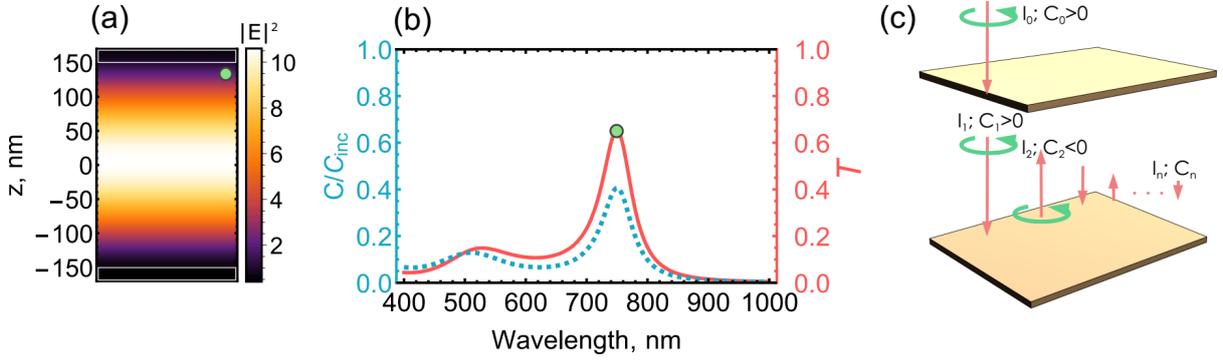

**Fig. 3.** (a) Electric field enhancement inside an empty 300 nm thick cavity induced by a normally incident plane wave at a wavelength of 750 nm corresponding to the first order Fabry-Pérot resonance. (b) Spectrum of normalized chirality density $C/C_{CPL}$ induced by a normally incident CP wave inside a 300 nm thick Fabry-Pérot cavity formed by 20 nm thick Au mirrors. (c) Illustration of chirality density limitation inside isotropic planar cavities: an incident CP plane wave experiences a series of reflections inside the cavity, reversing its handedness each time.

Preserving the helicity of reflected waves is crucial to overcoming this fundamental limitation with planar structures [30]. A possible way to beat the free space limit is to employ anisotropic mirrors with at most $C_z^2$ symmetry. Broken isotropy of the mirrors will allow to convert the polarization rotation direction upon reflection (CCW to CW and vice versa), thus preserving the sign of the chirality density of the travelling wave and enhancing the chirality density in the middle of the cavity [31]. Another recently suggested approach to bypass this limitation is based on oblique-propagating modes, which do not flip chirality upon reflection from specially designed metasurface mirrors [32]. Potentially, both of these approaches could allow obtaining enhanced CD and chiral mode splitting at the same time.

**FDTD simulations: test of a real system**

In order to test the analytical predictions, we perform a proof-of-principle experiment and the corresponding FDTD simulations with a coupled chiral nanoparticles-cavity system (see Appendix B for details of simulations). To ensure strong interaction of the chiral scatterer with the cavity, we will use chiral plasmonic nanocrescents[33–35] instead of molecules due to their large oscillator strength. Fig. 4(a) shows the top, tilted, and side view of the nanocrescent model used in the simulations. This shape is not only chiral, but is also highly anisotropic. We shall consider the effect of this anisotropy below. Absorption cross-section spectra of a single Au nanocrescent in vacuum upon illumination with CP light demonstrate a series of prominent peaks at ~600, ~700 nm, and ~1150 nm, respectively, Fig. 4(b) (see Fig. S1 for scattering cross-section spectra). In the following we will be focusing on the resonances lying in the visible range. Optical chirality of our nanoparticle is further corroborated by multipole decomposition (see Appendix C) of a single nanocrescent placed in an electric anti-node of a linearly polarized standing wave, Fig. 4(c), which shows excitation of a magnetic dipole moment collinear with the induced electric dipole moment at the resonance wavelength of ~700 nm.

The simulated electric field distributions in the horizontal plane excited by two circular polarizations at the 700 nm resonance reveal similar spatial distributions, Fig. 4(d), suggesting that the same eigenmode is excited with two orthogonal polarizations. Interestingly, this is in contrast to bigger dielectric nanocrescents of the same shape, where excitation with opposite circular polarizations reveals quite different field distributions and more complex eigenmodes structure [33].

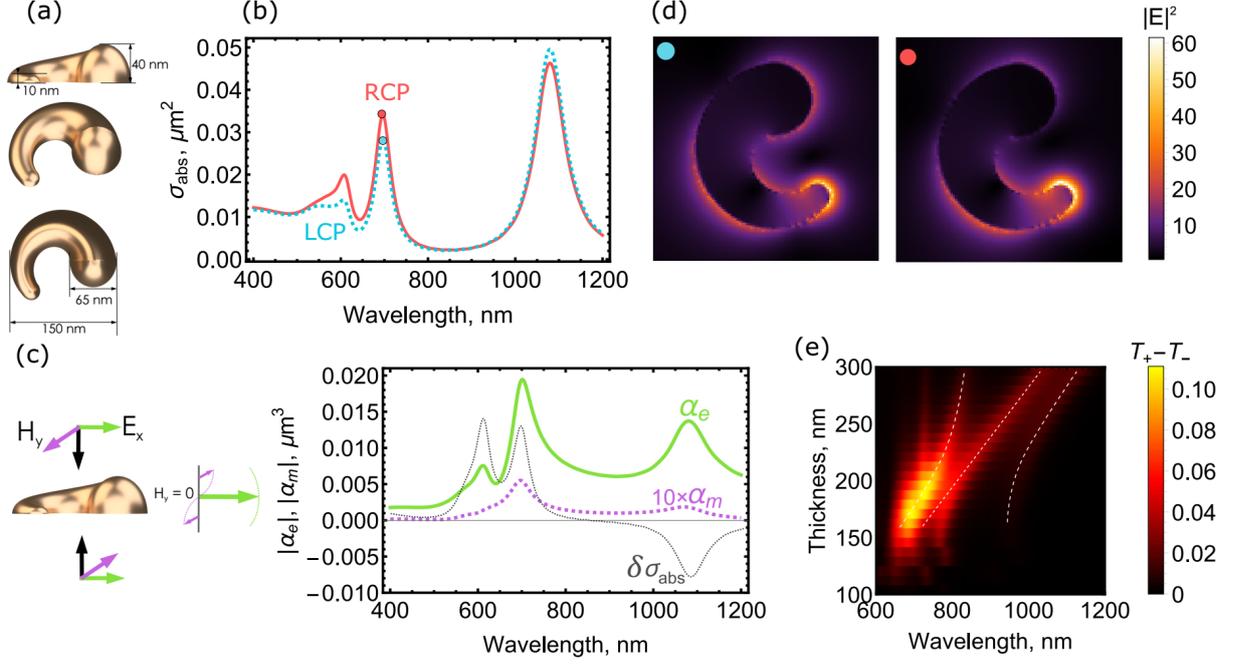

**Fig. 4**. FDTD simulations of the chiral nanocrescent. (a) Side, tilted, and top views of the Au nanocrescent model used in FDTD simulations rendered with the use of 3D software. (b) Simulated absorption cross-section spectra of a single Au nanocrescent in vacuum illuminated by a right- and left-circularly polarized plane wave from top. (c) Multipole decomposition of a single Au nanocrescent in vacuum placed in an electric field anti-node of a standing wave with $x-$polarized electric field (see inset). Solid curves depict the $x-$components of induced electric and magnetic dipoles, converted to electric and magneto-electric polarizability units ($\mu m^3$) for convenient normalization; dashed curve shows the differential absorption cross-section $\sigma_{abs}^{RCP} - \sigma_{abs}^{LCP}$. (d) Simulated electric field intensity distributions excited by two circular polarizations in the horizontal plane at the base of a single Au nanocrescents in air. (e) Map of the simulated CD spectra of the coupled plasmon-cavity system versus the cavity thickness at normal incidence for a square nanocrescent array with 300 nm period placed in the middle of the PMMA-filled cavity. The dashed lines are guides for the eye.

Next, we inspect the coupled plasmon-cavity system comprised by a square array of such nanocrescents with 300 nm period positioned in the middle of a Fabry-Pérot cavity formed by two 20 nm gold mirrors and filled with PMMA ($n = 1.45$). CD spectra at normal incidence calculated as a function of the cavity thickness expectedly exhibit the predicted mode anti-crossing, Fig. 3(e). However, in addition to that, one can see a middle peak appearing in the CD spectra with almost linear dispersion following that of the empty cavity mode (see Fig. S2), which is not predicted by the TMM calculations.

In order to unravel the origin of the middle peak in transmission spectra, we must take the anisotropy of the nanocrescents into account. Non-equal total transmission, i.e., total intensity of transmitted light irrespective of its polarization state $T_+ = |t_{++}|^2 + |t_{-+}|^2$, $T_- = |t_{--}|^2 + |t_{+-}|^2$ (where $t_{ij}$ are the Jones transmission matrix elements in the basis of circular polarizations) of incident circularly polarized light through a planar structure, $T_+ \neq T_-$, can often be misinterpreted as a manifestation of chirality, whereas the structure is in fact not chiral[36]. The lack of the three-fold rotational $C_z^3$ symmetry allows conversion between two circular polarized states ($t_{+-} \neq 0, t_{-+} \neq 0$), ref. [36]. This polarization conversion is the source of non-equal total transmission $T_+ \neq T_-$, which has nothing to do with chirality of the structure. In order to distinguish this so called elliptical dichroism from true circular dichroism, one needs to calculate complex-valued Jones matrix elements: non-equality $t_{++} \neq t_{--}$ will point out to chirality of the system. Calculated spectra of transmission Jones matrix elements for an array of nanocrescents in vacuum (see Fig. S3) indicate that the observed response of Au nanocrescents originates from both their chirality *and* polarization conversion.

Polarization conversion in transmission is not the only consequence of the violated $C_z^3$ symmetry of the nanoparticles. We argue that it also enables the middle peak in transmission and CD spectra of the coupled system. Eigenmodes of the Au nanocrescent participating in the coupling with the cavity modes have elliptical, rather than circular polarization due to its violated $C_z^3$ symmetry and induced generalized anisotropy[36]. Therefore, an incident circular polarization can be decomposed into the corresponding elliptical part, which will excite the coupled system and exhibit the mode splitting, and the remaining orthogonal part, which will not interact with the particles and show up in the transmission (or absorption) spectra close to the energy of the uncoupled cavity.

To substantiate this argument, we perform control FDTD simulations with a square array of Au gammadions, which are characterized by $C_z^4$ rotational symmetry (so that the resulting array is characterized by the *p4* wallpaper symmetry group), placed in the middle of the cavity, and the same gammadions in which one central bar has been removed (with the array being characterized by the *p2* wallpaper group). While $C_z^4$ symmetric particles exhibit a clear mode splitting and anti-crossing in transmission under illumination with circularly polarized light, Fig. 5(a), the broken $C_z^4$ symmetry particles exhibit the bright middle peak in the transmission following the uncoupled cavity dispersion, Fig. 5(b). Correspondingly, the *p4* symmetric array shows nearly zero differential transmission spectra $T_+ - T_-$ (with non-zero values attributed to the numerical errors of FDTD simulations), while the *p2* symmetric array clearly shows non-zero differential transmission caused by polarization conversion (see Fig. S4). Note that both systems chosen for this demonstration are not geometrically chiral due to preserved $xy$-plane of mirror symmetry.

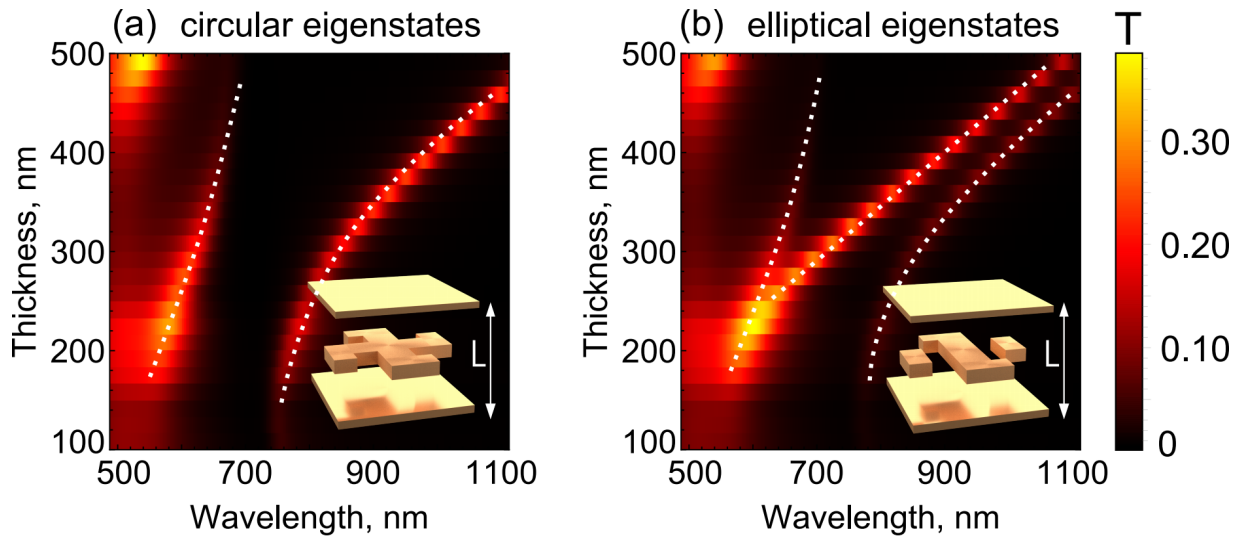

**Fig. 5**. Effect of particle anisotropy on the mode splitting and anti-crossing under illumination with circularly polarized light. (a) Transmission spectra through an array of Au gammadions with 125 nm period placed inside the middle of an empty Fabry-Pérot cavity formed by two 20 nm thick Au mirrors as a function of the cavity thickness. Dashed lines are guides for the eye marking approximate dispersion of polariton modes. (b) The same as (a) but for an array of gammadions with removed central bar.

**Experiment**

Lastly, we carry out a proof-of-principle experiment to confirm the prediction of mode splitting in CD of a chiral strongly coupled system. To that end, we have fabricated chiral gold nanocrescents using hole-mask colloidal lithography using techniques described in refs. [33,35] (see Appendix D for further details). Fig. 6a shows an SEM image of the fabricated gold nanocrescents on glass substrate, which indicates that the nanocrescents are distributed rather homogeneously over the large area (see Fig. S5). Unpolarized transmission spectra through bare particles on glass substrate were then measured (see Appendix E for further details), revealing a broad transmission dip around 800 nm, see Fig. 6(b). This broad dip can be attributed to overlapping resonances of particles with varying sizes and/or aspect ratios, resulting in a shift of the plasmon resonance. Correspondingly, the CD spectrum of the bare particles (calculated as the differential total transmission $T_+ - T_-$) shows a broad peak around 800 nm, Fig. 6(c).

Observed redshift of the plasmon peak from 700 nm in FDTD simulations to 800 nm in experiment is likely caused by the effect of the glass substrate.

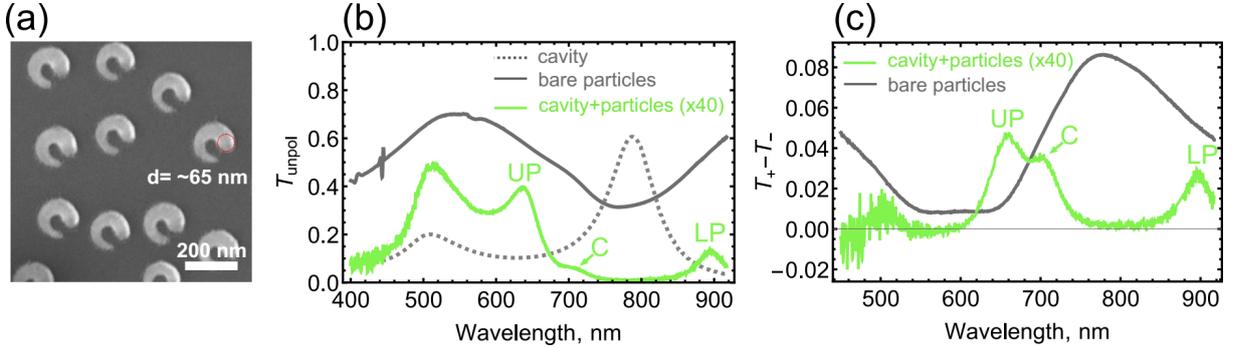

**Fig. 6**. Experiments with chiral hybrid cavity-plasmon system. (a) SEM image of chiral Au nanocrescents on glass substrate with ~65 nm head diameter, assumed for FDTD simulations. (b) Unpolarized ($\frac{T_+ + T_-}{2}$) transmission spectra of bare chiral nanocrescents on glass substrate, bare Fabry-Pérot cavity, and that of the coupled cavity-plasmon system. UP and LP label the position of upper and lower cavity-plasmon polaritons, whereas C labels the middle "uncoupled" peak due to in-plane anisotropy of the particles. (c) CD spectra of the same two systems, showing emerging mode splitting in CD of the coupled plasmon-cavity structure.

To arrange a coupled system, gold nanocrescents were deposited on top of half-cavity consisting of 20 nm gold film (serving as a bottom mirror) on glass substrate and ~100 nm thick $SiO_2$ layer, using same technique as bare nanoparticles. Then, the half-cavity with nanoparticles was completed by depositing 100 nm of PMMA layer, and another 20 nm gold film as a top mirror. Thickness of the FP cavity was ~200 nm to tune cavity resonance around ~800 nm (the dashed curve in Fig. 6b).

Unpolarized transmission spectrum of the coupled system at normal incidence shows the presence of two peaks at around 620 and 900 nm in Fig. 6b, unambiguously indicating emergence of the mode splitting due to strong coupling of plasmonic oscillators to the cavity mode [28]. Transmission spectrum also reveals a middle peak at around 700 nm, which originates from the nanocrescents anisotropy, as discussed above. Finally, the resulting CD spectra also clearly demonstrate splitting of the initial plasmon CD peak into two hybrid cavity-particle modes located at the nearly same wavelengths of 650 and 900 nm, as shown in Fig. 6c, thus confirming our theoretical prediction. We stress one more time that the observed asymmetric transmission originates from both chirality and anisotropy of the plasmonic nanostructure.

Transmission and CD spectra measured from strongly coupled systems with different geometrical parameters presented in Fig. S6 demonstrate qualitatively similar behavior.

The demonstrated the effect of CD mode splitting is reminiscent of the CD enhancement from chiral molecules by plasmon resonances demonstrated by Govorov and colleagues [11,12,14]. In their studies, the plasmon resonance of a metallic nanoparticle in the visible range was interacting with the tail of the chiral electronic transition of a molecule at higher energies. The plasmon resonance picks up the tail of the chiral molecule in the visible range and amplifies it due to the excitation of the superchiral near fields. This scenario can equally be interpreted as the mode splitting of CD with the uncoupled components being largely detuned in the first place, whereas we achieve the splitting by means of a large coupling constant.

**Conclusion**

In conclusion, we have demonstrated that generalized mode splitting can be induced in circular dichroism spectra of a chiral system, such as a molecule of a nanoparticle scatterer, by strong coupling to an achiral optical cavity. The resulting polariton modes inherit the chiral nature of its constituents and exhibit the mode splitting – the hallmark of the strong coupling regime – not only in scattering and absorption, but also in their CD spectra, thus allowing to split dichroic response of the system. At the same time, the use of planar cavities for strong coupling imposes fundamental limitations on the enhancement of CD spectra, which originate from the nature of circularly polarized standing waves. The predicted effect was observed in a proof-of-principle experiment involving chiral plasmonic nanoparticles coupled to a Fabry-Pérot cavity.

**Acknowledgements**
We thank Constantin Simovski for helpful discussion. The work was supported by the Swedish Research Council Infrastructure grant (VR Miljö) and the Knut and Alice Wallenberg Foundation.

**Appendix A: transfer-matrix method.** We adopt the formalism of $T$-matrix, describing propagation of plane waves in a planar multilayer structure [37]. The multilayer system is broken down into two types of primitive elements: homogeneous slabs and heterogeneous interfaces between two different media. For each element, the respective $T$-matrix connects complex amplitudes of forward- and backward propagating waves on the left and right sides of the element:

$$\begin{pmatrix} c \\ d \end{pmatrix} = \begin{pmatrix} T_{11} & T_{12} \\ T_{21} & T_{22} \end{pmatrix} \begin{pmatrix} a \\ b \end{pmatrix},$$

where a and b are amplitudes of the forward and backward wave on the left side, and c and d are those on the right side, respectively. In an isotropic chiral medium the propagating eigenmodes are circularly polarized plane waves [29]; furthermore, the rotation direction of CP waves is preserved (although the handedness reverses upon reflection). Therefore, the T-matrix will relate amplitudes of waves having the same rotation direction.

For a homogeneous layer of thickness $L$ the $T$-matrix is diagonal (since there is no reflection in a homogeneous layer) and reads $T_i = \begin{pmatrix} e^{ik_{i,+}L} & 0 \\ 0 & e^{-ik_{i,-}L} \end{pmatrix}$, where $k_{i,+}$ and $k_{i,-}$ are wave vectors of forward and backward propagating waves in $i$-th layer, respectively. For a bi-isotropic reciprocal chiral media the wave vector read [29] $k_{i,\pm} = \frac{\omega}{c}(\sqrt{\varepsilon\mu} \pm \kappa_i)$.

The $T$-matrix for an interface $T_{ij}$ can be derived from the corresponding scattering matrix that relates outgoing waves to the incoming ones: $\begin{pmatrix} b \\ c \end{pmatrix} = \begin{pmatrix} r_{ij} & t_{ji} \\ t_{ij} & r_{ji} \end{pmatrix} \begin{pmatrix} a \\ d \end{pmatrix}$, where $r_{ij}$ and $t_{ij}$ are the complex reflection and transmission electric field coefficients for incidence from medium $i$ toward medium $j$. For an interface between two bi-isotropic reciprocal chiral media, these coefficients read [29]: $r_{ij} = (\eta_{i,+}^{-1} - \eta_{j,+}^{-1})/(\eta_{i,-}^{-1} + \eta_{j,+}^{-1})$, $t_{ij} = 1 + r_{ij}$, where $\eta_{i,\pm} = \sqrt{\mu_i/\varepsilon_i}$ are characteristic impedances of the two media, and + and – denote forward and backward propagating waves (with respect to the incident wave, but not with respect to the global propagation direction of the problem). Resolving the interface scattering matrix with respect to c and d, we find $\begin{pmatrix} c \\ d \end{pmatrix} = \frac{1}{t_{ji}} \begin{pmatrix} t_{ij}t_{ji} - r_{ij}r_{ji} & r_{ji} \\ -r_{ij} & 1 \end{pmatrix} \begin{pmatrix} a \\ b \end{pmatrix}$.

The resulting T matrix of the entire system $T_\Sigma$ is found as the product of T matrices of consecutive interfaces and layers: $T_\Sigma = T_{n-1,n} \cdot T_n \cdot ... \cdot T_2 \cdot T_{12}$. Finally, resolving this matrix with respect to b and c and letting $d = 0$ (no incidence from the right side), we find the desired reflection and transmission coefficients:

$$t = T_{11}^{\Sigma} - \frac{T_{12}^{\Sigma} T_{21}^{\Sigma}}{T_{22}^{\Sigma}}, \quad r = -\frac{T_{21}^{\Sigma}}{T_{22}^{\Sigma}}.$$

**Appendix B: FDTD simulations.** Numerical simulations of the electromagnetic response of Au nanocrescents were performed using the finite-difference time-domain (FDTD) method using commercial software (Lumerical, USA). Scattering and absorption spectra, field distributions and multipole decomposition of a single particle were obtained with the use of the total field/scattered field (TFSF) source and PML boundary conditions. Transmission spectra of uncoupled and coupled cavities were obtained using a normal incidence plane wave source and periodic boundary conditions.

**Appendix C: multipole decomposition.** The Cartesian electric, toroidal, magnetic, and magnetic toroidal dipole moments of a single nanocrescent were calculated using the standard expansion formulas [38]:

$$\mathbf{P} = \varepsilon_0 (\varepsilon - 1) \int \mathbf{E} dV,$$

$$\mathbf{T} = \frac{k^2}{10} \varepsilon_0 (\varepsilon - 1) \int [(\mathbf{r} \cdot \mathbf{E})\mathbf{r} - 2r^2 \mathbf{E}] dV,$$

$$\mathbf{M} = -i \frac{\omega}{2} \varepsilon_0 (\varepsilon - 1) \int \mathbf{r} \times \mathbf{E} dV,$$

$$\mathbf{M}_t = i \frac{k^2 \omega}{20} \varepsilon_0 (\varepsilon - 1) \int r^2 \mathbf{r} \times \mathbf{E} dV.$$

The corresponding electric and magneto-electric dipole polarizabilities (their x-components) in $\mu m^3$ were then calculated as $\alpha_{ee} = (P_x + T_x)/(\varepsilon_0 E_{inc})$ and $\alpha_{em} = (M_x + M_{t,x})/(\varepsilon_0 E_{inc})$ with $E_{inc}$ being the electric field of the incident plane wave.

**Appendix D: sample fabrication.** Glass substrates were cleaned with hot acetone and isopropanol, respectively, to remove impurities. To fabricate coupled system, first, 20 nm of gold thin film for bottom mirror was prepared using an e-beam evaporator. Then, ~100-107.5 nm of SiO$_2$ layers were deposited using STS plasma-enhanced chemical vapor deposition (PECVD) at 300ºC, to form half-cavities. Later, chiral nanoparticles were deposited on top of the half-cavities or directly on bare glass substrates, to fabricate coupled systems or reference bare particles, respectively.

Chiral nanoparticles were fabricated using hole-colloidal lithography technique (HCL), using techniques described in refs. [33,35]. Firstly, ~240 nm thick poly(methyl methacrylate) (PMMA) layer was spin coated on the substrates (half-cavities or bare glass), baked at 180°C for 10 mins, and then functionalized with polydiallyldimethylammonium chloride (PDDA, 0.2%) aqueous solution for 30 seconds. Subsequently, 80 or 100 nm of polystyrene (PS) beads were applied on the functionalized-substrates. After 3 mins, the samples were rinsed gently with water and dried with nitrogen. Then, 10 nm of gold thin film was evaporated, and PS beads with 10 nm gold caps were removed with tape-stripping. Later, the samples were dry etched using oxygen plasma (4 mins, 50 W, 250 mTorr) to form hole-mask. For the deposition of chiral gold nanoparticles, 120 nm of gold was evaporated with rate of 2 Å/s at substrate tilt angle of 13° while rotating the substrate at a constant speed over total angle of 270°, followed by removal of PMMA resists with hot acetone and rinsed with isopropanol. Afterwards, the nanoparticles were annealed at 250°C for 10 min to improve the crystallinity of the particles.

For the coupled systems, PMMA layer with same thicknesses as the bottom $SiO_2$ half-cavity was spin coated and annealed at 180°C for 10 mins, followed by evaporation of 20 nm gold thin film as a top mirror to complete the cavity. Bare FP cavity was prepared with same procedures as coupled system, except of the nanoparticles deposition.

**Appendix E: optical characterization.** Polarization-resolved transmission spectra were measured using circularly polarized white light from a halogen light as a broadband illumination. Broadband right or left polarized light were obtained by transmitting the white light through a polarizer misaligned of either +45° or -45° with respect to a Fresnel rhomb, respectively [33]. Transmission spectra of the samples were collected using a fiber coupled spectrometer (BWTek). Reference spectrum was taken with bare glass substrate under same measurement condition.